\documentclass[12pt]{article}
\usepackage{amsmath}
\usepackage{amsfonts}
\usepackage{amsthm}
\usepackage{graphicx}
\newtheorem{thm}{Theorem}[section]

\newtheorem{lem}[thm]{Lemma}

\newtheorem{rem}[thm]{Remark}

\begin{document}

\begin{center}
{\Large Characteristic Lie Algebra and Classification of
Semi-Discrete Models}

\vskip 0.2cm

{Ismagil Habibullin}\footnote{e-mail: habibullin\_i@mail.rb.ru,
(On leave from Ufa Institute of Mathematics, Russian Academy of
Science, Chernyshevskii Str. , 112, Ufa, 450077, Russia)}

{Asl{\i} Pekcan}\footnote{e-mail: asli@fen.bilkent.edu.tr}

{Department of Mathematics, Faculty of Science,
 \\Bilkent University, 06800, Ankara, Turkey \\}

\end{center}
\vskip 0.2cm
\begin{abstract}
Characteristic Lie algebras of semi-discrete chains are studied.
The attempt to adopt this notion to the classification of Darboux
integrable chains has been undertaken.
\end{abstract}

\section{Introduction}

Investigation of the class of hyperbolic type differential
equations of the form
\begin{equation}
u_{xy}=f(x,y,u,u_x,u_y) \label{hyp}
\end{equation}
has a very long history. Various approaches have been developed to
look for particular and general solutions of these kind equations.
In the literature one can find several definitions of
integrability of the equation. According to one given by G.
Darboux, equation (\ref{hyp}) is called integrable if it is
reduced to a pair of ordinary (generally nonlinear) differential
equations, or more exactly, if its any solution satisfies the
equations of the form \cite{Darboux}, (see also
\cite{GrundlandVassiliou})
\begin{equation}
F(x,y,u,u_x,u_{xx},..., D_x^mu)=a(x),\quad G(x,y,u,u_y,u_{yy},...,
D_y^nu)=b(y),\label{ch}
\end{equation}
for appropriately chosen functional parameters $a(x)$ and $b(y)$,
where $D_x$ and $D_y$ are operators of differentiation with
respect to $x$ and $y$, $u_x=D_xu$, $u_{xx}=D_xu_x$ and so on.
Functions $F$ and $G$ are called $x$- and $y$-integrals of the
equation respectively.

An effective criterion of Darboux integrability has been proposed
by G. Darboux himself. Equation (\ref{hyp}) is integrable if and
only if the Laplace sequence of the linearized equation terminates
at both ends. A rigorous proof of this statement has been found
only recently \cite{SokolovZhiber}, \cite{AndersonKamran}.

 An alternative method of investigation and classification of the
Darboux integrable equations has been developed by A. B. Shabat
based on the notion of characteristic Lie algebra. Let us give a
brief explanation of this notion. Begin with the basic property of
the integrals. Evidently each $y$-integral satisfies the
condition: $D_yF(x,y,u,u_x,u_{xx},..., D_x^mu)=0$. Taking the
derivative by applying the chain rule one defines a vector field
$X_1$ such that
\begin{equation}\label{X1}
X_1F=\Big(\frac{\partial}{\partial y}+ u_y\frac{\partial}{\partial
u}+ f\frac{\partial}{\partial u_x}+D_x(f)\frac{\partial}{\partial
u_{xx}}+...\Big)F=0.
\end{equation}
So the vector field $X_1$ solves the equation $X_1F=0$. But in
general, the coefficients of the vector field depend on the
variable $u_y$ while the solution $F$ does not. This puts a severe
restriction on $F$, actually $F$ should satisfy one more equation
$X_2F=0$, where $X_2=\frac{\partial}{\partial u_y}$. Now the
commutator of these two operators will also annulate $F$.
Moreover, for any $X$ from the Lie algebra generated by $X_1$ and
$X_2$ one gets $XF=0$. This Lie algebra is called characteristic
Lie algebra of the equation (\ref{hyp}) in the direction of $y$.
Characteristic algebra in the $x$-direction is defined in a
similar way. Now by virtue of the famous Jacobi theorem, equation
(\ref{hyp}) is Darboux integrable if and only if both of its
characteristic algebras are of finite dimension. In
\cite{ShabatYamilov} and \cite{LeznovSmirnovShabat}, the
characteristic Lie algebras for the systems of nonlinear
hyperbolic equations and their applications are studied.

In this article we will study semi-discrete chains of the
following form
\begin{equation}\label{dhyp}
t_{1x}=f(t,t_1,t_x)
\end{equation}
from the Darboux integrability point of view. Here the unknown
$t=t(n,x)$ is a function of two independent variables: one
discrete $n$ and one continuous $x$. It is assumed that
$\frac{\partial f}{\partial t_x}\neq 0$. Subindex means shift or
derivative, for instance, $t_1=t(n+1,x)$ and
$t_x=\frac{\partial}{\partial x}t(n,x)$. Below we use $D$ to
denote the shift operator and $D_x$ to denote the $x$-derivative:
$Dh(n,x)=h(n+1,x)$ and $D_xh(n,x)=\frac{\partial}{\partial
x}h(n,x)$. For the iterated shifts use the subindex $D^jh=h_j$.

Introduce now notions of the integrals for the semi-discrete chain
(\ref{dhyp}). The $x$-integral is defined similar to the
continuous case. We call a function $F=F(x,n,t,t_1,t_2,...)$
depending on a finite number of shifts $x$-integral of the chain
(\ref{dhyp}), if the following condition is valid $D_xF=0$. It is
natural, in accordance with the continuous case, to call a
function $I=I(x,n,t,t_x,t_{xx},...)$ $n$-integral of the chain
(\ref{dhyp}) if it is in the kernel of the difference operator:
$(D-1)I=0$. In other words $n$-integral should still unchanged
under the action of the shift operator $DI=I$, (see also
\cite{AdlerStartsev}). One can write it in an enlarged form
\begin{equation}\label{I}
I(x,n+1,t_1,f,f_x,f_{xx},...)=I(x,n,t,t_x,t_{xx},...).
\end{equation}
Notice that it is a functional equation, the unknown is taken at
two different "points". This circumstance causes the main
difficulty in studying discrete chains. Such kind problems appear
when one tries to apply the symmetry approach to discrete
equations (see \cite{NijhoffCapel}, \cite{GKP}). However the
concept of the Lie algebra of characteristic vector fields
provides an effective tool to investigate chains.

Introduce vector fields in the following way. Concentrate on the
main equation (\ref{I}). Evidently the left hand side of it
contains the variable $t_1$ while the right hand side does not.
Hence the total derivative of the function $DI$ with respect to
$t_1$ should vanish. In other words the $n$-integral is in the
kernel of the operator $Y_1:=D^{-1}\frac{\partial}{\partial
t_1}D$. Similarly one can check that $I$ is in the kernel of the
operator $Y_2:=D^{-2}\frac{\partial}{\partial t_1}D^2$. Really,
the right hand side of the equation $D^2I=I$ which immediately
follows from (\ref{I}) does not depend on $t_1$, therefore the
derivative of the function $D^2I$ with respect to $t_1$ vanishes.
Proceeding this way one can easily prove that for any natural $j$
the operator $Y_j=D^{-j}\frac{\partial}{\partial t_1}D^j$ solves
the equation $Y_jI=0$.

So far we shifted the argument $n$ forward, shift it now backward
and use the main equation (\ref{I}) written as $D^{-1}I=I$.
Rewrite the original equation (\ref{dhyp}) in the form
\begin{equation}\label{dhypg}
t_{-1x}=g(t,t_{-1},t_x).
\end{equation}
This can be done because of the condition $\frac{\partial
f}{\partial t_x}\neq0$ assumed above. In the enlarged form the
equation $D^{-1}I=I$ looks like
\begin{equation}\label{Ig}
I(x,n-1,t_{-1},g,g_x,g_{xx},...)=I(x,n,t,t_x,t_{xx},...).
\end{equation}
The right of the last equation does not depend on $t_{-1}$ so the
total derivative of $D^{-1}I$ with respect to $t_{-1}$ is zero,
i.e. the operator $Y_{-1}:=D\frac{\partial }{\partial t_{-1}}
D^{-1}$ solves the equation $Y_{-1}I=0$. Moreover, the operators
$Y_{-j}=D^{j}\frac{\partial}{\partial t_{-1}}D^{-j}$ also satisfy
similar conditions $Y_{-j}I=0$.

Summarizing the reasonings above one can conclude that the
$n$-integral is annulated by any operator from the Lie algebra
$\tilde{L}_n$ generated by the operators \cite{Habibullin}
\begin{equation}\label{gen}
...,Y_{-2},Y_{-1},Y_{-0},Y_{0},Y_{1},Y_{2},...
\end{equation}
where $Y_0=\frac{\partial }{\partial t_{1}}$ and
$Y_{-0}=\frac{\partial }{\partial t_{-1}}$. The algebra
$\tilde{L}_n$ consists of the operators from the sequence
(\ref{gen}), all possible commutators, and linear combinations
with coefficients depending on the variables $n$ and $x$.
Evidently equation (\ref{dhyp}) admits a nontrivial $n$-integral
only if the dimension of the algebra $\tilde{L}_n$ is finite. But
it is not clear that the finiteness of dimension $\tilde{L}_n$
enough for existence of $n$-integrals. By this reason we introduce
another Lie algebra called the characteristic Lie algebra of the
equation (\ref{dhyp}). First we define in addition to the
operators $Y_1, Y_2,...$ differential operators
$X_j=\frac{\partial}{\partial_{t_{-j}}}$ for $j=1,2,...$\,.\\

The following theorem allows us to define the characteristic Lie
algebra.
\begin{thm}\label{thm1}
Equation (\ref{dhyp}) admits a nontrivial $n$-integral if and only
if the following two conditions hold:\\
1)  Linear envelope of the operators $\{Y_j \}_1^{\infty}$ is
of finite dimension, denote this dimension $N$;\\
2)  Lie algebra $L_n$ generated by the operators
${Y_1,Y_2,...,Y_N,X_1,X_2,...,X_N}$ is of finite dimension. We
call $L_n$ the characteristic Lie algebra of (\ref{dhyp}).
\end{thm}
\begin{rem}\label{remark}
It is easy to prove that if dimension of $\{Y_j \}_1^{\infty}$ is
$N$ then the set $\{Y_j \}_1^{N}$ constitute a basis in the linear
envelope of $\{Y_j \}_1^{\infty}$.
\end{rem}
\section{Characteristic Lie Algebra $L_n$}

\noindent Study some properties of the characteristic Lie algebra
introduced in the theorem \ref{thm1} above. Begin with the proof
of the remark \ref{remark}. It will immediately follow from the
following lemma.
\begin{lem}
If for some integer $N$ the operator $Y_{N+1}$ is a linear
combination of the operators with less indices:
\begin{equation}\label{YM}
Y_{N+1}=\alpha_1Y_1+\alpha_2 Y_2+...+\alpha_NY_N
\end{equation}
then for any integer $j>N$, we have a similar expression
\begin{equation}
Y_{j}=\beta_1Y_1+\beta_2 Y_2+...+\beta_NY_N.
\end{equation}
\end{lem}
\noindent \textbf{Proof.} Due to the property
$Y_{k+1}=D^{-1}Y_kD$, we have from (\ref{YM})
\begin{equation}
Y_{N+2}=D^{-1}(\alpha_1)Y_2+D^{-1}(\alpha_2)Y_3+...+D^{-1}(\alpha_N)(\alpha_1Y_1+...+\alpha_NY_N).
\end{equation}
Now by using induction one can easily complete the proof of the
lemma.
\begin{lem} The following commutativity relations take place:
$$ [Y_{0},Y_{-0}]=0,\quad [Y_{0},Y_{1}]=0, \quad [Y_{-0},Y_{-1}]=0.\quad $$
\end{lem}

\noindent {\bf Proof.} The first of the relations is evident. In
order to prove two others find the coordinate representation of
the operators $Y_{1}$ and $Y_{-1}$ acting in the class of locally
smooth functions of the variables $x,n,t,t_x,t_{xx},...$\,. By
direct computations
\begin{eqnarray}\label{Y1I}
Y_1 I&=&D^{-1} \displaystyle \frac{d}{dt_1} D I\nonumber\\\nonumber\\
&=&D^{-1} \displaystyle \frac{d}{dt_1}
I(t_1,f,f_x,...)\nonumber\\\nonumber\\
&=&\Big\{\frac{\partial}{\partial t}+D^{-1}\Big(\frac{\partial
f}{\partial t_1}\Big)\frac{\partial}{\partial
t_x}+D^{-1}\Big(\frac{\partial f_x}{\partial
t_1}\Big)\frac{\partial}{\partial
t_{xx}}+...\Big\}I(t,t_x,t_{xx},...)\nonumber\\\nonumber\\
\end{eqnarray}
one gets
\begin{eqnarray}\label{Y1}
Y_1=\frac{\partial}{\partial t}+D^{-1}\Big(\frac{\partial
f}{\partial t_1}\Big)\frac{\partial}{\partial
t_x}+D^{-1}\Big(\frac{\partial f_x}{\partial
t_1}\Big)\frac{\partial}{\partial
t_{xx}}+D^{-1}\Big(\frac{\partial f_{xx}}{\partial
t_1}\Big)\frac{\partial}{\partial t_{xxx}}+...\,.
\end{eqnarray}
Now notice that all of the functions  $f$, $f_x$, $f_{xx}, ...$
depend on the variables $t_1,t,t_x,t_{xx}, ...$ and do not depend
on $t_2$ hence the coefficients of the vector field $Y_1$ do not
depend on $t_1$ and therefore the operators $Y_1$ and $Y_0$
commute. In a similar way by using the explicit coordinate
representation
\begin{eqnarray}\label{Y-1}
Y_{-1}=\frac{\partial}{\partial t}+D\Big(\frac{\partial
g}{\partial t_{-1}}\Big)\frac{\partial}{\partial
t_x}+D\Big(\frac{\partial g_x}{\partial
t_{-1}}\Big)\frac{\partial}{\partial t_{xx}}+D\Big(\frac{\partial
g_{xx}}{\partial t_{-1}}\Big)\frac{\partial}{\partial
t_{xxx}}+...\,.
\end{eqnarray}
one can prove that $[Y_{-0},Y_{-1}]=0.$

The following statement turned out to be very useful for studying
the characteristic Lie algebra $L_n$.
\begin{lem} Suppose that the vector field
\begin{equation}
Y=\alpha(0)\partial_t+\alpha(1)\partial_{t_x}+\alpha(2)\partial_{t_{xx}}+...,
\end{equation}
where $\alpha_x(0)=0$ solves the equation $[D_x,Y]=0$, then
$Y=\alpha(0)\partial_t$.
\end{lem}
The proof is based on the following formula
\begin{equation}\label{DxY}
[D_x,Y]=(\alpha_x(0)-\alpha(1))\partial_t+(\alpha_x(1)-\alpha(2))\partial_{t_x}+...\,.
\end{equation}
So if $a_x(0)=0$, then $a(1)=0$, but if $a_x(1)=0$ then $a(2)=0$
and hence $a(j)=0$ for all $j>0$.

In the formula (\ref{Y1I}) we have already given an enlarged
coordinate form of the operator $Y_1$. One can check that the
operator $Y_2$ is a vector field of the form
\begin{equation}\label{Y2}
Y_2=D^{-1}(Y_1(f))\partial_{t_x}+D^{-1}(Y_1(f_x))\partial_{t_{xx}}
+D^{-1}(Y_1(f_{xx}))\partial_{t_{xxx}}+...\,.
\end{equation}
It immediately follows from the equation $Y_2=D^{-1}Y_1D$ and the
coordinate representation (\ref{Y1I}). By induction one can prove
similar formulas for arbitrary $j$:
\begin{equation}
Y_{j+1}=D^{-1}(Y_j(f))\partial_{t_x}+D^{-1}(Y_j(f_x))\partial_{t_{xx}}
+D^{-1}(Y_j(f_{xx}))\partial_{t_{xxx}}+...\,.
\end{equation}
\begin{lem}\label{lemma24} For the operators $D_x$, $Y_1$, $Y_{-1}$
considered on the space of smooth functions of $t,t_x,t_{xx},...$
the following commutativity relations take place:
$$[D_x,Y_1] = pY_1 \quad, \quad [D_x,Y_{-1}] = qY_{-1},$$
\noindent where $p=- D^{-1} (\frac{\partial f}{\partial t_1})$ and
$q=- D (\frac{\partial g}{\partial t_{-1}})$.
\end{lem}
\noindent \textbf{Proof.} Recall that
\begin{equation}
Y_1=\frac{\partial}{\partial t}+D^{-1}\Big(\frac{\partial
f}{\partial t_1}\Big)\frac{\partial}{\partial
t_x}+D^{-1}\Big(\frac{\partial f_x}{\partial
t_1}\Big)\frac{\partial}{\partial t_{xx}}+...\,.
\end{equation}
We find $[D_x,Y_1]$ by using (\ref{DxY}) as
\begin{equation}\label{DxY1}
[D_x,Y_1]=-D^{-1}(f_{t_1})\partial_t+D^{-1}(D_x(f_{t_1})-f_{xt_1})\partial_{t_x}+...\,.
\end{equation}
\noindent For arbitrary function $H$, we have
\begin{eqnarray}
[D_x,\partial_{t_1}]H(t,t_1,t_x,t_{xx},...)&=&D_xH_{t_1}-\frac{\partial}{\partial_{t_1}}D_xH\nonumber\\
&=&(H_{tt_1}t_x+H_{t_1t_1}{t_1}_x+...)-\frac{\partial}{\partial_{t_1}}(H_tt_x+H_{t_1}{t_1}_x+...)\nonumber\\
&=&-H_{t_1}f_{t_1}.
\end{eqnarray}
\noindent By taking $H=f$ and $H=f_x$ one gets
$[D_x,\partial_{t_1}]f=-f_{t_1}f_{t_1}$,
$[D_x,\partial_{t_1}]f_x=-f_{xt_1}f_{t_1},$ and so on. We insert
these equations into (\ref{DxY1}) to find
\begin{eqnarray}
[D_x,Y_1]&=&- D^{-1} \Big(\displaystyle \frac{\partial
 f}{\partial t_1}\Big)\Big\{\frac{\partial}{\partial
t}+D^{-1}\Big(\frac{\partial f}{\partial
t_1}\Big)\frac{\partial}{\partial t_x}+D^{-1}\Big(\frac{\partial
f_x}{\partial t_1}\Big)\frac{\partial}{\partial t_x}+...\Big\}\nonumber\\
&=&- D^{-1} \Big(\displaystyle \frac{\partial
 f}{\partial t_1}\Big)Y_1.
\end{eqnarray}
In a similar way one can prove that $[D_x,Y_{-1}]=-
D\Big(\displaystyle \frac{\partial
 g}{\partial t_{-1}}\Big)Y_{-1}$.

Let us prove theorem $1$.$1$. Suppose that there exists a
nontrivial $n$-integral $F=F(t,t_x,...,t_{[N]})$ for the equation
(\ref{dhyp}), here $t_{[j]}=D_x^jt$ for any $j\geq 0$. Then all
the vector fields from the Lie algebra $M$ generated by $\{Y_j,X_k
\}$ for $j=1,2,...$ and $k=1,...,N_2$ where $N_2$ is chosen
arbitrarily satisfying $N_2\geq N$ annulate $F$. We will show that
dim $M$ $< \infty$. Consider first the projection of the algebra
$M$ given by the operator $P_N$:
\begin{equation}\label{projection}
 P_{N}\Big(\displaystyle \sum_{i=-N_2}^{-1}x(i)\partial_{t_i}+\displaystyle \sum_{i=0}^{\infty}
 x(i)\partial_{t_{[i]}}\Big)=\displaystyle
 \sum_{i=-N_2}^{-1}x(i)\partial_{t_i}+\displaystyle \sum_{i=0}^{N}x(i)\partial_{t_{[i]}}.
 \end{equation}
Let $L_n(N)$ be the projection of $M$. Then evidently the equation
$Z_0F=0$ is satisfied for any $Z_0$ in $L_n(N)$. Evidently, dim
$L_n(N) < \infty$. Let the set $\{Z_{01},Z_{02},...,Z_{0N_1}\}$
forms a basis in $L_n(N)$. One can represent any $Z_0$ in $L_n(N)$
as a linear combination
\begin{equation}\label{Z0}
Z_0=\alpha_1Z_{01}+\alpha_2Z_{02}+...+\alpha_{N_1}Z_{0N_1}.
\end{equation}
Suppose that the vector fields $Z,Z_1,...,Z_{N_1}$ in $M$ are
connected with the operators $Z_0,Z_{01},...,Z_{0N_1}$ in $L_n(N)$
by the formulas $P_N(Z)=Z_0,
P_N(Z_1)=Z_{01},...,P_N(Z_{N_1})=Z_{0N_1}$. We have to prove that
\begin{equation}\label{proofZ}
Z=\alpha_1Z_1+\alpha_2Z_2+...+\alpha_{N_1}Z_{N_1}.
\end{equation}
In the proof, we use the following lemma.
\begin{lem}
Let $F_1=D_xF$ and $F$ is an $n$-integral. Then for each $Z$ in
$M$ we have $ZF_1=0$.
\end{lem}
\noindent \textbf{Proof.} It is easy to check that $F_1$ is also
an $n$-integral, really $DF_1=DD_xF=D_xDF=D_xF=F_1$. It was shown
above that any $Z$ in $M$ annulates $n$-integrals.

Apply the operator $Z-\alpha_1 Z_1-\alpha_2
Z_2-...-\alpha_{N_1}Z_{N_1}$ to the function
$F_1=F_1(t,t_x,t_{xx},...,t_{[N+1]})$,
\begin{equation}\label{ZF1}
(Z-\alpha_1Z_1-\alpha_2Z_2-...-\alpha_{N_1}Z_{N_1})F_1=0.
\end{equation}
We can write (\ref{ZF1}) as
\begin{eqnarray}\label{longZ0}
&&(Z_0-\alpha_1Z_{01}-\alpha_2Z_{02}-...-\alpha_{N_1}Z_{0N_1}) F_1
+(X(N+1)-\alpha_1X_1(N+1)\nonumber\\&-&
\alpha_2X_2(N+1)-...-\alpha_{N_1}X_{N_1}(N+1))\frac{\partial}{\partial
t_{{[N+1]}}}F_1=0,\nonumber\\
&&
\end{eqnarray}
where $X(N+1), X_1(N+1),...,X_{N_1}(N+1)$ are the coefficients
before $\partial_{t_{[N+1]}}$ of the vector fields
$Z,Z_1,Z_2,...,Z_{N_1}$. The first summand in (\ref{longZ0})
vanishes by (\ref{Z0}). In the second one the factor
$\frac{\partial}{\partial t_{[N+1]}} F_1=\frac{\partial}{\partial
t_{[N]}}F$ is not zero. So we have
\begin{equation}\label{XN+1}
X(N+1)=\alpha_1X_1(N+1)+\alpha_2X_2(N+1)+...+\alpha_{N_1}X_{N_1}(N+1).
\end{equation}
Equation (\ref{XN+1}) shows that
\begin{equation}
P_{N+1}(Z)=\alpha_1P_{N+1}(Z_1)+\alpha_2P_{N+1}(Z_2)+...+\alpha_{N_1}P_{N+1}(Z_{N_1}).
\end{equation}
So by applying mathematical induction, one can prove the formula
(\ref{proofZ}). Thus the Lie algebra $M$ is of finite dimension.
Now construct the characteristic algebra $L_n$ by using $M$. Since
dim $M$ $< \infty$, the linear envelope of the vector fields
$\{Y_j \}_1^{\infty}$ is of finite dimension. Choose a basis in
this linear space consisting of $Y_1,Y_2,...,Y_K$ for $K\leq N
\leq N_2$. Then the algebra generated by
$Y_1,Y_2,...,Y_K,X_1,X_2,...,X_K$ is of finite dimension, because
it is a subalgebra of $M$. This algebra is just characteristic Lie
algebra of the equation (\ref{dhyp}).

Suppose that conditions $(1)$ and $(2)$ of the theorem \ref{thm1}
are satisfied. So there exists a finite dimensional characteristic
Lie algebra $L_n$ for the equation (\ref{dhyp}). Show that in this
case equation (\ref{dhyp}) admits a nontrivial $n$-integral. Let
$N_1$ is the dimension of $L_n$ and $N$ is the dimension of the
linear envelope of the vector fields $\{Y_j \}_{j=1}^{\infty}$.
Take the projection $L_n(N_2)$ of the Lie algebra $L_n$ defined by
the operator $P_{N_2}$ defined by the formula
(\ref{projection})with $N_2$ instead of $N$. Evidently, $L_n(N_2)$
consists of the finite sums $Z_0=\displaystyle
\sum_{i=-N}^{-1}x(i)\partial_{t_i}+\displaystyle
\sum_{i=0}^{N_2}x(i)\partial_{t_{[i]}}$ where $N=N_1-N_2$. Let
$Z_{01},...,Z_{0N_1}$ form a basis in $L_n(N_2)$. Then we have
$N_1=N+N_2$ equations $Z_{0j}G=0$, $j=1,...,N_1$, for a function
$G$ depending on $N+N_2+1=N_1+1$ independent variables. Then due
to the well-known Jacobi theorem, there exists a function
$G=G(t_{-N},t_{-N+1},...,t_{-1},t,t_x,t_{xx},...,t_{[N_2]})$,
which satisfies the equation $ZG=0$ for any $Z$ in $L_n$. But
really it does not depend on $t_{-N}$, ..., $t_{-1}$ because
$X_1G=0$, $X_2G=0$, ...., $X_{N}G=0$. Thus the function $G$ is
$G=G(t,t_x,t_{xx},...,t_{[N_2]})$. Such a function is not unique
but any other solution of these equations, depending on the same
set of the variables, can be represented as $h(G)$ for some
function $h$.

Note one more property of the algebra $L_n$. Let $\pi$ be a map
which assigns to each $Z$ in $L_n$ its conjugation $D^{-1}ZD$.
Evidently, the map $\pi$ acts from the algebra $L_n$ into its
central extension $L_n \oplus \{X_{N+1}\}$, because for the
generators of $L_n$ we have $D^{-1}Y_jD=Y_{j+1}$ and
$D^{-1}X_jD=X_{j+1}$. Evidently, $[X_{N+1},Y_j]=0$ and
$[X_{N+1},X_j]=0$ for any integer $j\leq N$. Moreover $X_{N+1}F=0$
for the function $G=G(t,t_x,...,t_{[N_2]})$ mentioned above. This
fact implies that $ZG_1=0$ for $G_1=DG$ and for any vector field
$Z$ in $L_n$. Really, for any $Z$ in $L_n$ one has a
representation of the form $D^{-1}ZD=\tilde{Z}+\lambda X_{N+1}$
where $\tilde{Z}$ in $L_n$ and $\lambda$ is a function. So
\begin{equation}
ZG_1=ZDG=D(D^{-1}ZDG)=D(\tilde{Z}+\lambda X_{N+1})G=0.
\end{equation}
Therefore $G_1=h(G)$ or $DG=h(G)$. In other words function
$G=G(n)$ satisfies an ordinary difference equation of the first
order. Its general solution can be written as $G=H(n,c)$ where $H$
is a function of two variables and $c$ is an arbitrary constant.
By solving the equation $G=H(n,c)$ with respect to $c$ one gets
$c=F(G,n)$. The function $F=F(G,n)$ found is just $n$-integral
searched. Actually, $DF(G,n)=Dc=c=F(G,n)$. So $DF=F$. This
completes the proof of the theorem $1$.$1$.

\section{Restricted classification}

Different approaches are known to classify integrable nonlinear
differential (pseudo-differential) equations. One of the most
popular and powerful is based on higher symmetries. For the first
time the theoretical aspects of this method have been formulated
in the famous paper by N. Kh. Ibragimov and A. B. Shabat
\cite{IbragimovShabat}. Several classes of nonlinear models were
tested by this method in \cite{MSY}, \cite{LeviYamilov}. The
symmetry approach allowed R. I. Yamilov to find all integrable
chains of the Volterra type \cite{Yamilov}:
$u_t(n)=f(u(n-1),u(n),u(n+1))$. The consistency approach to the
classification of integrable discrete equations has been studied
by Adler, Bobenko and Suris in \cite{AdlerBobenkoSuris}.
Classification based on the notion of the recursion operator is
studied in \cite{Gurses1}-\cite{Gurses4}.

In this paper we undertake an attempt to adopt the notion of
characteristic Lie algebra to the problem of classification of
Darboux integrable discrete equations of the form (\ref{dhyp}).
The classification problem consists of describing of all chains
admitting finite dimensional characteristic Lie algebras in both
directions. Actually the problem of studying the algebra generated
by the operators (\ref{gen}) seems to be quite difficult. That is
why we will start with a very simple case.

{\bf Formulation of the problem.} Study the problem of finding all
of the equations (\ref{dhyp}) for which the Lie algebra generated
by the operators $Y_1$ and $Y_{-1}$ is two dimensional.

Let us denote $Y_{1,-1} =[Y_1,Y_{-1}]$. We require that the
following relation takes place $Y_{1,-1}=\lambda Y_1+\mu Y_{-1}$.
It follows from the explicit formulas (\ref{Y1}) and (\ref{Y-1})
that the vector field $Y_{1,-1}$ does not contain a summand with
the term $\frac{\partial}{\partial t}$ hence $\mu=-\lambda$. The
commutators of the basic vector fields with the operator of the
total derivative admit simple expressions (see lemma 2.4 above).
Evaluate the commutator $[D_x,Y_{1,-1}]$
\begin{eqnarray*}
[D_x,Y_{1,-1}]&=&[Y_1,[D_x,Y_{-1}]]-[Y_{-1},[D_x,Y_{1}]]\\
&=&[Y_1,qY_{-1}]-[Y_{-1},pY_1]\\
&=&Y_1(q)Y_{-1}+qY_{1,-1}-Y_{-1}(p)Y_1+pY_{1,-1}\\
&=&(p+q)Y_{1,-1}+Y_1(q)Y_{-1}-Y_{-1}(p)Y_1.
\end{eqnarray*}
Remind that due to the reasoning above a coefficient
$\lambda=\lambda (n,x)$ should exist such that
\begin{equation}\label{Y1-1}
Y_{1,-1}=\lambda (Y_1-Y_{-1}).
\end{equation}
The problem is in finding of $f$ in the equation
${t_{1}}_x=f(t,t_1,t_x)$ for which the constraint (\ref{Y1-1}) is
valid.

Commute each side of the equation (\ref{Y1-1}) with the operator
$D_x$
\begin{eqnarray*}
[D_x,Y_{1,-1}]&=&[D_x,\lambda Y_1]-[D_x,\lambda Y_{-1}]\\
&=&(p+q)\lambda (Y_1-Y_{-1})+Y_1(q)Y_{-1}-Y_{-1}(p)Y_1\\
&=&D_x(\lambda)Y_1+\lambda p Y_1-D_x(\lambda) Y_{-1}-\lambda q
Y_{-1}.
\end{eqnarray*}
Compare now two different expressions for the commutator. This
gives rise to conditions
$$q\lambda-Y_{-1}(p)=D_x(\lambda) \quad , \quad
p\lambda -Y_1(q)=D_x(\lambda),$$ \noindent which form an
over-determined system of equations for unknown $\lambda$, it
should satisfy simultaneously two equations. By solving them with
respect to $\lambda$ and $D_x(\lambda)$ we obtain equations
\begin{equation}
\displaystyle \lambda=\frac{Y_{-1}(p)-Y_1(q)}{q-p}\quad , \quad
D_x(\lambda)=\frac{qY_1(q)-pY_{-1}(p)}{p-q}
\end{equation}
which immediately yield
\begin{equation}\label{conditioncharalgebra}
\displaystyle
D_x\Big(\frac{Y_{-1}(p)-Y_1(q)}{q-p}\Big)=\frac{pY_{-1}(p)-qY_1(q)}
{q-p}.
\end{equation}
First of all note that this equation contains both $f$ and its
inverse $g$. Exclude $g$.  Recall that ${t_1}_x=f(t,t_1,t_x)$ and
$t_x=f(t_{-1},t,t_{-1x})$ where ${t_{-1x}}=g(t,t_{-1},t_x)$.
Differentiating the identity $t_x=f(t_{-1},t,g(t,t_{-1},t_x))$
with respect to $t_{-1}$ one gets
\begin{equation}
\displaystyle D^{-1}\Big(\frac{\partial f}{\partial
t}(t,t_1,t_x)\Big)+D^{-1}\Big(\frac{\partial f}{\partial
t_x}(t,t_1,t_x)\Big)\frac{\partial g}{\partial t_{-1}}=0
\end{equation}
which implies that
\begin{equation}
g_{t_{-1}}=-D^{-1}\Big(\frac{f_t}{f_{t_x}} \Big),
\end{equation}
so $\displaystyle D(g_{t_{-1}})=-\frac{f_t}{f_{t_x}}$. Let us
write the equation (\ref{conditioncharalgebra}) explicitly.
Evaluate first $Y_1(q)$ and $Y_{-1}(p)$, where
$p=-D^{-1}(f_{t_1})$ and $q=\frac{f_t}{f_{t_x}}$.
\begin{eqnarray}
Y_1(q)&=&\Big\{\partial_t+D^{-1}(f_{t_1})\partial_{t_x}+D^{-1}(f_{xt_1})\partial_{t_{xx}}+...\Big\}
\frac{f_t}{f_{t_x}}\nonumber\\
&=&\Big(\frac{f_t}{f_{t_x}}\Big)_t
+D^{-1}(f_{t_1})\Big(\frac{f_{t}}{f_{t_x}}\Big)_{t_x}.
\end{eqnarray}
\begin{eqnarray}
Y_{-1}(p)&=&-\Big\{\partial_t-\frac{f_t}{f_{t_x}}\partial_{t_x}
-D\Big(\frac{\partial g_x}{\partial t_{-1}}\Big)\partial_{t_{xx}}-...\Big\}D^{-1}(f_{t_1})\nonumber\\
&=&-\Big(D^{-1}(f_{t_1})\Big)_t+\frac{f_t}{f_{t_x}}
\Big(D^{-1}(f_{t_1})\Big)_{t_x}.
\end{eqnarray}
By inserting these equations into (\ref{conditioncharalgebra}) one
gets a long expression
\begin{eqnarray}\label{longeq}
\displaystyle
&&D_x\Bigg\{\frac{-\Big(D^{-1}(f_{t_1})\Big)_t+\frac{f_t}{f_{t_x}}
\Big(D^{-1}(f_{t_1})\Big)_{t_x}
-\Big(\Big(\frac{f_t}{f_{t_x}}\Big)_t
+D^{-1}(f_{t_1})\Big(\frac{f_{t}}{f_{t_x}}\Big)_{t_x}\Big)}
{\frac{f_t}{f_{t_x}}+D^{-1}(f_{t_1})}\Bigg\}
\nonumber\\
&=&\frac{D^{-1}(f_{t_1})\Big(\Big(D^{-1}(f_{t_1})\Big)_t-\frac{f_t}{f_{t_x}}
\Big(D^{-1}(f_{t_1})\Big)_{t_x}\Big)
-\frac{f_t}{f_{t_x}}\Big(\Big(\frac{f_t}{f_{t_x}}\Big)_t
+D^{-1}(f_{t_1})\Big(\frac{f_{t}}{f_{t_x}}\Big)_{t_x}\Big)}
{\frac{f_t}{f_{t_x}}+D^{-1}(f_{t_1})}.\nonumber\\ &&
\end{eqnarray}
Equation (\ref{longeq}) is rather difficult to study and we put
one more restriction on $f$. Suppose that $f=a(t)+b(t_1)+c(t_x)$.
Find the variables $p$, $q$, $Y_1(q)$, $Y_{-1}(p)$ in terms of
$a$, $b$, $c$
\begin{eqnarray*}
&& \displaystyle p=-D^{-1}(f_{t_1})=-b'(t),\\\\
&& \displaystyle
q=-D(g_{t_{-1}})=\frac{f_t}{f_{t_x}}=\frac{a'(t)}{c'(t_x)},\\\\
&&Y_1(q)=(\partial_t+b'(t)\partial_{t_x})\frac{a'(t)}{c(t_x)}
=\frac{a''(t)}{c'(t_x)}-\frac{b'(t)a'(t)c''(t_x)}{(c'(t_x))^2},\\\\
&&Y_{-1}(p)=\Big(\partial_t-\frac{a'(t)}{c'(t_x)}\partial_{t_x}\Big)(-b'(t))=-b''(t).
\end{eqnarray*}
Substitute these expressions into (\ref{conditioncharalgebra})
\begin{equation}\label{COND}
D_xG(t,t_x)=\displaystyle \frac{b'(t)b''(t)-\displaystyle
\frac{a'(t)}{c'(t_x)}\Bigg[\displaystyle \frac{a''(t)}{c'(t_x)}-
\frac{b'(t)a'(t)c''(t_x)}{(c'(t_x))^2}\Bigg]} {\displaystyle
\frac{a'(t)}{c'(t_x)}+b'(t)}
\end{equation}
where,
\begin{equation}
G(t,t_x)=\Bigg[ \frac{-b''(t)- \displaystyle
\frac{a''(t)}{c'(t_x)}+\displaystyle
\frac{b'(t)a'(t)c''(t_x)}{(c'(t_x))^2} } {\displaystyle
\frac{a'(t)}{c'(t_x)}+b'(t)} \Bigg].
\end{equation}
Evidently the left hand side of the equation (\ref{COND}) is of
the form $\frac{\partial G}{\partial t}t_x+\frac{\partial
G}{\partial t_x}t_{xx}$ and it contains the variable $t_{xx}$
while the right hand side does not contain it. This gives an
additional constraint $\frac{\partial G}{\partial t_x}=0$.

Investigation of the equation (\ref{COND}) is tediously long. Thus
we will only give the answers.
\begin{thm}\label{lastthm}
If the equation (\ref{dhyp}) with particular choice of
$f(t,t_1,t_x)=a(t)+b(t_1)+c(t_x)$ has the operators $Y_1$ and
$Y_{-1}$ such that the Lie algebra generated by these two
operators is two-dimensional then $f(t,t_1,t_x)$ is of one of the
forms:\\

\noindent
$1)$ $f(t,t_1,t_x)=\gamma(t_x+\beta \sinh (\alpha t+\lambda))+\beta \cosh t_1+\eta$,\\
$2)$ $f(t,t_1,t_x)=\gamma(t_x+\beta e^{\alpha t})+\beta e^{\alpha t_1}+\eta$,\\
$3)$ $f(t,t_1,t_x)=c(t_x)+\gamma t_1+\beta$,\\
$4)$ $ f(t,t_1,t_x)=\gamma \ln |t_x|+\displaystyle \frac{1}{\gamma}\ln (e^t-e)+\beta$,\\
$5)$ $f(t,t_1,t_x)=\gamma{t_x}^2+\beta{t_x}+\alpha t+\eta$,\\

\noindent where $c(t_x)$ is an arbitrary function and $\alpha$,
$\beta$, $\gamma$, $\lambda$, $\eta$ are arbitrary constants.

\end{thm}

 \noindent Additionally, if in the cases $1)$, $2)$, $3)$, the
corresponding characteristic Lie algebras are also two-dimensional
then the equations are of the forms:\\
$i)$ $t_{1x}=t_x+\beta \sinh(t)+\beta \cosh(t_1)$,\\
$ii)$ $t_{1x}=t_x+e^t+e^{t_1}$,\\
$iii)$ $t_{1x}=t_x$,\\
\noindent and they have the $n$-integrals respectively:\\
$a)$ $I=\frac{\beta^2}{2}\cosh^2 t -\beta t_x \cosh
t+\frac{t_x^2}{2}+\beta t_x \sinh t-t_{xx}+\frac{\beta^2}{2}n$,\\
$b)$ $I=\frac{e^{2t}}{2}+\frac{t_x^2}{2}-t_{xx}$,\\
$c)$ $I=t_{xx}$.\\
\noindent Note that $ii)$ has also $x$-integral as
\begin{equation}
F=e^{t_1-t}+e^{2t_1-t_2-t}+e^{t_1-t_2}.
\end{equation}
Thus, $t_{1x}=t_x+e^t+e^{t_1}$ is a discrete analog of the
Liouville equation.
 The proof of the theorem \ref{lastthm} is
given in the next section.

\newpage

\section{The Proof of the Theorem \ref{lastthm}}
\noindent Here we consider the cases which satisfy the condition
(\ref{COND}) and prove the theorem \ref{lastthm}.\\

\noindent \textbf{Case 1.} $a'=0$:\\

\noindent The condition (\ref{COND}) turns out to be
\begin{equation}
D_x\Big( \frac{-b''}{b'}\Big)=b''.
\end{equation}
This gives us $b'=\gamma$ so $b(t_1)=\gamma t_1+\beta$. There is
no condition on $c(t_x)$. Hence $f$ becomes
\begin{equation}
f(t,t_1,t_x)=c(t_x)+\gamma t_1+\beta,
\end{equation}
where $\beta$ and $\gamma$ are arbitrary constants and $c(t_x)$ is
an arbitrary function of $t_x$.\\

\noindent \textbf{Case 2.} $b'=0$:\\

\noindent The condition (\ref{COND}) turns out to be
\begin{eqnarray*}
D_x\Big( \frac{a''}{a'}\Big)&=&\Big(\frac{a'''a'-(a'')^2}{(a')^2}\Big)t_x\\
&=&\frac{a''}{c'}.
\end{eqnarray*}
Since only $c$ depends on $t_x$, we have $\displaystyle
c'=\frac{\gamma}{t_x}$ which implies
\begin{equation}
c(t_x)=\gamma \ln |t_x|+\beta.
\end{equation}
\noindent The remaining equation gives $a(t)$ as
\begin{equation}
a(t)=-\ln (-1+e^{\lambda t+\lambda \sigma}+\xi).
\end{equation}
Thus, in a simplified form $f$ is
\begin{equation}
f(t,t_1,t_x)=\gamma \ln |t_x|+\displaystyle \frac{1}{\gamma}\ln
(e^t-e)+\beta,
\end{equation}
where $\beta$ and $\gamma$ are arbitrary constants.\\

\noindent \textbf{Case 3.} $c'''=0$:\\

\noindent Note that we should have
\begin{equation}\label{Gt_x}
\displaystyle \frac{\partial G}{\partial_{t_x}}(t,t_x)=0
\end{equation}
\noindent in the condition (\ref{COND}) since there is no $t_x$
explicitly in the right hand side. If we expand $(\ref{Gt_x})$ we
obtain
\begin{equation}
c'''c'[b'(a')^2+c'(b')^2a']-(c'')^2[b'(a')^2+2c'(b')^2a']+(c')^2c''[a''b'-a'b'']=0.
\end{equation}
\noindent Under the case $c'''=0$ we have $c''=A$ and naturally
$c'=At_x+B$ where $A$, $B$ are constants, we have
\begin{eqnarray*}
A^2{t_x}^2[a''b'-a'b'']&+&2At_x[B(a''b'-a'b'')-A(b')^2a']
\\&+&[-Ab'(a')^2-2AB(b')^2a'+B^2a''b'-B^2a'b'']=0.
\end{eqnarray*}
\noindent The coefficient of ${t_x}^2$ should vanish, so
$a''b'-a'b''=0$. We should also have
\begin{equation}\label{eqn}
[B(a''b'-a'b'')-A(b')^2a']=0
\end{equation}
\noindent from the coefficient of $t_x$. Since $a''b'-a'b''=0$,
the equation becomes
\begin{equation}
A(b')^2a'=0.
\end{equation}
\noindent To provide this equation to be satisfied we have three
choices.\\

\noindent \textbf{Subcase 3.i.} $A=0$:\\

\noindent In this case $c''=0$ and so $c'=B$ where $B$ is
constant. The condition (\ref{COND}) becomes
\begin{equation}\label{subcase1}
D_x\Big[\displaystyle \frac{a''+Bb''}{a'+Bb'}\Big]=\displaystyle
\frac{\displaystyle \frac{a'a''}{B}-Bbb''}{a'+Bb'}.
\end{equation}
\noindent Hence
\begin{equation}
\displaystyle \frac{a''+Bb''}{a'+Bb'}=\alpha,
\end{equation}
\noindent where $\alpha$ is constant. Thus we have two unknowns
and two equations which are
\begin{equation}
a'=\sigma e^{\alpha t}-Bb',
\end{equation}
\noindent and
\begin{equation}
(a')^2=B^2(b')^2+\rho,
\end{equation}
\noindent where $\sigma$ and $\rho$ are constants. We take square
of the first equation and subtract form the second one. We get
\begin{equation}\label{eqforb}
\sigma^2e^{2\alpha t}-2\sigma Bb'e^{\alpha t}-\rho=0
\end{equation}
\noindent whose solution is
\begin{equation}
b= \displaystyle \frac{\sqrt{\rho}}{B\alpha}\cosh (\alpha t+\log
\displaystyle \frac{\sigma}{\sqrt{\rho}}) + \delta.
\end{equation}
\noindent In a similar way we find $a$ as
\begin{equation}
a=\displaystyle \frac{\sqrt{\rho}}{B\alpha}\sinh (\alpha t+\log
\displaystyle \frac{\sigma}{\sqrt{\rho}}) + \tau.
\end{equation}
\noindent Here $\rho \neq 0$ and $\delta$, $\tau$ are constants.
Since $c'=B$, we find $c(t_x)=Bt_x+\zeta$. Finally, after
appropriate transformation we obtain
\begin{equation}
f(t,t_1,t_x)=\gamma(t_x+\beta\sinh (\alpha t+\lambda))+\beta \cosh
t_1+\eta,
\end{equation}
\noindent where $\alpha$, $\beta$, $\eta$, $\gamma$ and $\lambda$
are
constants.\\

\noindent If $\rho=0$, then the equation (\ref{eqforb}) gives us
\begin{equation}
b=\displaystyle \frac{\sigma}{2B\alpha}e^{\alpha t}.
\end{equation}
\noindent By using this equation we obtain $a$ as
\begin{equation}
a=\displaystyle \frac{\gamma}{2\alpha}e^{\alpha t}.
\end{equation}
\noindent We have also $c(t_x)=Bt_x+\xi$. Thus simplified form of
$f$ is
\begin{equation}
f(t,t_1,t_x)=\gamma(t_x+\beta e^{\alpha t})+\beta e^{\alpha
t_1}+\eta,
\end{equation}
\noindent where $\alpha$, $\beta$, $\gamma$ and $\eta$ are
arbitrary constants.
\newpage

\noindent \textbf{Subcase 3.ii.} $a'=0$:\\

\noindent In the $\textbf{Case 1.}$ we have analyzed this case.
Here,
\begin{eqnarray*}
f(t,t_1,t_x)&=&a(t)+b(t_1)+c(t_x)\\
&=&\gamma {t_x}^2+\beta{t_x}+\alpha t_1+\eta,
\end{eqnarray*}
\noindent where $\alpha$, $\beta$, $\gamma$ and $\eta$ are arbitrary constants.\\

\noindent \textbf{Subcase 3.iii.} $b'=0$:\\

\noindent Here $c(t_x)=\gamma {t_x}^2+\beta{t_x}+\lambda$, but for
$b'=0$ we have the following equation
\begin{eqnarray*}
D_x\Big( \frac{a''}{a'}\Big)&=&\Big(\frac{a'''a'-(a'')^2}{(a')^2}\Big)t_x\\
&=&\frac{a''}{c'}.
\end{eqnarray*}
which cannot be satisfied by this $c(t_x)$. Hence $a''=0$. So we
obtain,
\begin{eqnarray*}
f(t,t_1,t_x)&=&a(t)+b(t_1)+c(t_x)\\
&=&\gamma {t_x}^2+\beta {t_x}+\alpha t+\eta,
\end{eqnarray*}
\noindent where $\alpha$, $\beta$, $\gamma$ and $\eta$ are arbitrary constants.\\

\noindent The functions given in these cases may not give us
$n$-integral $I$ even though they satisfy the condition
(\ref{COND}). Let us see an example to this fact. This example
belongs to
$\textbf{Case 1.}$\\

\noindent \textbf{Example.}
\begin{equation}
t_{1x}=f(t,t_1,t_x)=t_1+t_x.
\end{equation}
\noindent By using the definitions of the operators $Y_j$, we
obtain

\begin{eqnarray*}
Y_1&=&\partial_t+\partial_{t_x}+\partial_{t_{xx}}+\partial_{t_{xxx}}+...\\\\
Y_2&=&\partial_{t_x}+2\partial_{t_{xx}}+3\partial_{t_{xxx}}+...\\\\
Y_3&=&\partial_{t_x}+3\partial_{t_{xx}}+6\partial_{t_{xxx}}+...
\end{eqnarray*}

\noindent and so on. As we see that the set
${Y_1,Y_2,Y_3,...,Y_n}$ is linearly independent which means that
the algebra is infinite dimensional. By the theorem \ref{thm1},
$n$-integral does not exist.

Now we will focus on three problems which yield from the above
considerations. We will study on $n$-integrals of these
problems.\\

\noindent \textbf{Problem 1.} The equation for the function $f$ is
\begin{equation}\label{problem1f}
t_{1x}=f(t,t_1,t_x)=\lambda t_x+\lambda \beta \sinh t+\beta \cosh
t_1 +\gamma.
\end{equation}
Hence,
\begin{equation}\label{problem1g}
t_{-1x}=g(t,t_x,t_{-1})=\displaystyle \frac{1}{\lambda}
(t_x-\lambda \beta \sinh t_{-1}-\beta \cosh t-\gamma).
\end{equation}
\noindent To obtain the operators $Y_1$ we find the equations
written below:

\begin{eqnarray*}
f_{t_1}&=&\beta \sinh t_1,\\\\
f_x &=& \lambda t_{xx}+\lambda \beta t_x \cosh t+\beta t_{1x}\sinh
t_1,\\\\f_{xt_1}&=&\beta \cosh t_1 (\lambda t_x+\lambda \beta
\sinh t +\beta \cosh t_1 +\gamma)+\beta^2 \sinh^2t_1 ,\\\\
f_{xx}&=&\lambda t_{xxx}+\lambda \beta (t_x)^2 \sinh t+\lambda
\beta t_{xx}\cosh t+\beta (t_{1x})^2\cosh
      t_1\\\\
      &+&\beta \sinh t_1(\lambda t_{xx}
+\lambda \beta t_x\cosh t+\beta t_{1x}\sinh t_1),\\\\
f_{xxt_1}&=&3\beta^2 (t_{1x})\sinh t_1\cosh t_1+\beta
(t_{1x})^2\sinh t_1+\beta^3 \sinh^3 t_1\\\\
&+& \beta \cosh t_1(\lambda t_{xx}+\lambda \beta t_x\cosh t+\beta
t_{1x}\sinh t_1),
\end{eqnarray*}
\begin{eqnarray*}
f_{xxx}&=&\lambda t_{xxx}+\lambda \beta t_{xxx}\cosh t+3\lambda
\beta t_xt_{xx}\sinh t +\lambda \beta t_x^3\cosh t\\\\
&+&3\beta\cosh t_1 (t_1x)[ \lambda t_{xx}+\lambda \beta t_x \cosh
t+\beta
t_{1x}\sinh t_1]+\beta (t_{1x})^3\sinh t_1 \\\\
&+&\beta \sinh t_1 \{\lambda t_{xxx}+\lambda \beta t_{xx}\cosh
t+\lambda \beta t_x^2\sinh t+\beta(t_{1x}^2)\cosh t_1  \\\\
&+&[\lambda t_{xx}+\lambda \beta t_x\cosh t+\beta t_{1x}\sinh
t_1]\beta \sinh t_1\},\\\\
f_{xxxt_1}&=&3\beta[\lambda t_{xx}+\lambda \beta t_x\cosh t+\beta
t_{1x}\sinh t_1][\beta \sinh t_1\cosh t_1+t_{1x}\sinh t_1]\\\\
&+&3\beta t_{1x}\cosh t_1[\beta^2\sinh^2 t_1+\beta t_{1x} \cosh
t_1]+3\beta^2(t_{1x})^2\sinh^2t_1\\\\
&+&\beta \cosh t_1 [\lambda t_{xxx}+\lambda \beta t_{xx} \cosh
t+\lambda \beta t_x^2\sinh t+\beta t_{1x}^2 \cosh t_1\\\\&+&\beta
\sinh t_1(\lambda t_{xx}+\lambda \beta t_x\cosh t+\beta
t_{1x}\sinh t_1)]+\beta t_{1x}^3\cosh t_1\\\\
&+&\beta \sinh t_1[\beta \cosh t_1(\lambda t_{xx}+\lambda \beta
t_x\cosh t+\beta t_{1x}\sinh t_1)+\beta^3 \sinh^3 t_1\\\\
&+&3\beta^2 t_{1x}\cosh t_1\sinh t_1+\beta t_{1x}^2\sinh t_1],\\\\
&\vdots&
\end{eqnarray*}
Thus,
\begin{eqnarray*}
D^{-1}(f_{t_1})&=&\beta \sinh t,\\\\
D^{-1}(f_{xt_1})&=&\beta t_x \cosh t +\beta^2 \sinh^2 t,\\\\
D^{-1}(f_{xxt_1})&=&3\beta^2t_x\sinh t\cosh t+\beta t_x^2\sinh t
+\beta t_{xx}\cosh t+\beta^3 \sinh^3t ,
\end{eqnarray*}
\begin{eqnarray*}
D^{-1}(f_{xxxt_1})&=&4\beta^2 t_x^2 \sinh^2t+6\beta^3 t_x \cosh
t\sinh^2 t+4\beta^2t_{xx}\sinh t \cosh t\\\\&+&3\beta^2 t_x^2
\cosh^2 t+\beta t_x^3 \cosh t+3\beta t_xt_{xx}\sinh t\\\\
&+&\beta t_{xxx}\cosh t+\beta^4 \sinh^4t,\\\\
 &\vdots&
\end{eqnarray*}
\noindent Let us define $\beta \sinh t=\psi$. Hence we can write
the operator $Y_1$ as
\begin{eqnarray}
Y_1&=&\partial_t+\psi \partial_{t_{x}}
+(\psi^2+\psi_x)\partial_{t_{xx}}
+[(\psi^2+\psi_x)_x+\psi(\psi^2+\psi_x)]\partial_{t_{xxx}}\nonumber\\\nonumber\\
&+&\{[(\psi^2+\psi_x)_x+\psi(\psi^2+\psi_x)]_x+\psi[(\psi^2+\psi_x)_x+\psi(\psi^2+\psi_x)]
\}\partial_{t_{xxxx}}\nonumber\\\nonumber\\
&+&...\,.
\end{eqnarray}
\noindent Now we will obtain the operator $Y_{-1}$. From
(\ref{problem1g}) we have,
\begin{eqnarray*}
g_{t_{-1}}&=&-\beta \cosh t_{-1},\\\\
g_x &=& [t_{xx}-\lambda \beta t_{-1x}\cosh t_{-1}-\beta t_x\sinh t]/\lambda,\\\\
g_{xt_{-1}}&=&\beta^2 \cosh^2 t_{-1}-\beta t_{-1x} \sinh t_{-1},\\\\
g_{xx}&=&[t_{xxx} -\beta \cosh t_{-1}(t_{xx}-\lambda \beta
t_{-1x}\cosh t_{-1}-\beta t_x
\sinh t)\\\\
&-&\beta t_{xx}\sinh t-\beta t_x^2\cosh t-\lambda \beta
(t_{-1x})^2\sinh t_{-1}]/\lambda,\\\\
g_{xxt_{-1}}&=&[3\lambda \beta^2 (t_{-1x})\sinh t_{-1}\cosh
t_{-1}-\lambda \beta (t_{-1x})^2\cosh t_{-1}-\lambda \beta^3 \cosh^3 t_{-1}\\\\
&-&\beta \sinh t_{-1}(t_{xx}-\lambda \beta t_{-1x}\cosh
t_{-1}-\beta t_x \sinh t)]/\lambda,
\end{eqnarray*}
\begin{eqnarray*}
 g_{xxx}&=&\{t_{xxxx}+ -3\beta
t_{-1x}\sinh t_{-1}(t_{xx}-\lambda \beta t_{-1x}\cosh t_{-1}-\beta
t_x\sinh t)
\\\\&-&\lambda \beta (t_{-1x})^3\cosh t_{-1}
-\beta t_{xxx}\sinh t-3\beta t_xt_{xx}\cosh t-\beta t_x^3\sinh
t\\\\
&-&\beta \cosh t_{-1}[t_{xxx}-\beta \cosh t_{-1}(t_{xx}-\lambda
\beta t_{-1x}\cosh t_{-1}-\beta t_x \sinh t)\\\\&-& \lambda \beta
(t_{-1x})^2\sinh t_{-1}-\beta t_{xx}\sinh t-\beta t_x^2\cosh t]\}/\lambda,\\\\
g_{xxxt_{-1}}&=&\{(t_{xx}-\lambda \beta t_{-1x}\cosh t_{-1}-\beta
t_x \sinh t)(4\beta^2 \cosh t_{-1}\sinh t_{-1}-3\beta t_{-1x}\cosh
t_{-1})\\\\
&-&3\beta t_{-1x}\sinh t_{-1}(2\lambda \beta^2 \cosh^2
t_{-1}-\lambda \beta t_{-1x}\sinh t_{-1})\\\\
 &-&\beta \sinh t_{-1}[t_{xxx}-\beta \cosh
t_{-1}(t_{xx}-\lambda \beta t_{-1x}\cosh t_{-1}-\beta t_x \sinh
t)\\\\
&-& \lambda \beta (t_{-1x})^2\sinh t_{-1}-\lambda \beta
(t_{-1x})^3\sinh t_{-1}-\beta t_{xx}\sinh t-\beta t_x^2\cosh
t]\\\\
&+&\lambda \beta^4 \cosh^4 t_{-1}+4\lambda
\beta^2t_{-1x}^2\cosh^2 t_{-1} \}/\lambda,\\\\
&\vdots&
\end{eqnarray*}
From these equations we get
\begin{eqnarray*}
D(g_{t_{-1}})&=&-\beta \cosh t,\\\\
D(g_{xt_{-1}})&=&\beta^2 \cosh^2 t-\beta t_x \sinh t,\\\\
D(g_{xxt_{-1}})&=&3\beta^2 t_x \sinh t\cosh t-\beta t_x^2\cosh t
-\beta t_xx \sinh t-\beta^3 \cosh^3 t,
\end{eqnarray*}
\begin{eqnarray*}
D(g_{xxxt_{-1}})&=&4\beta^2t_{xx}\sinh t \cosh t-3\beta
t_xt_{xx}\cosh t-6\beta^3 t_x\cosh^2 t\sinh t\\\\&+&4\beta^2
t_x^2\cosh^2t+3\beta^2t_x^2\sinh^2t -\beta t_x^3\sinh t\\\\
&-&\beta t_{xxx}\sinh t+\beta^4 \cosh^4 t,\\\\
&\vdots&
\end{eqnarray*}
\noindent If we define $-\beta \cosh t=\varphi$ we write the
operator $Y_{-1}$ as

\begin{eqnarray}
Y_{-1}&=&\partial_t+\varphi
\partial_{t_x}+(\varphi^2+\varphi_x)\partial_{t_{xx}}+
[(\varphi^2+\varphi_x)_x+\varphi(\varphi^2+\varphi_x)]\partial_{t_{xxx}}\nonumber\\\nonumber\\
&+&\{[(\varphi^2+\varphi_x)_x+\varphi(\varphi^2+\varphi_x)]_x
+\varphi[(\varphi^2+\varphi_x)_x+\varphi(\varphi^2+\varphi_x)]\}\partial_{t_{xxxx}}\nonumber\\\nonumber\\
&+&...\,.
\end{eqnarray}
\noindent Now let us only take the first three terms of $Y_1$ and
$Y_{-1}$ and find $n$-integral $I$ such that both equations $Y_1
I=0$ and $Y_{-1}I=0$ are satisfied. The first three terms of $Y_1$
are
\begin{equation}
Y_1=\partial_t+\beta \sinh t
\partial_{t_x}+(\beta \cosh t_x+\beta^2 \sinh^2
t)\partial_{t_{xx}}.
\end{equation}
\noindent We will use the method of characteristic,
\begin{equation}
\displaystyle \frac{dt}{1}=\frac{dt_x}{\beta \sinh
t}=\frac{dt_{xx}}{(\beta \cosh t_x+\beta^2 \sinh^2
t)}=\frac{dI}{0}.
\end{equation}
\noindent By the first and second terms we have
\begin{equation}
\beta \cosh t-t_x=c_1.
\end{equation}
\noindent By the first and third terms we have
\begin{equation}
\beta t_x \sinh t-t_{xx}=c_2.
\end{equation}
\noindent Hence from $Y_1I=0$ we get the solution $I=F(c_1,c_2)$ where
$F$ is an arbitrary function. Now by using the transformation $\tilde{t}=t$,
$c_1=\beta \cosh t-t_x$, $c_2=\beta t_x \sinh t-t_{xx}$, we will write $Y_{-1}$
 in terms of $\tilde{t}$, $c_1$ and $c_2$. Note that
\begin{eqnarray*}
\displaystyle
\partial_t&=&\displaystyle \frac{\partial_{\tilde{t}}}{\partial_t}\frac{\partial}{\partial_{\tilde{t}}}
+\frac{\partial_{c_1}}{\partial_t}\frac{\partial}{\partial_{c_1}}
+\frac{\partial_{c_2}}{\partial_t}\frac{\partial}{\partial_{c_2}}=
\partial_{\tilde{t}}+\beta \sinh \tilde{t}\partial_{c_1}+\beta t_x \cosh \tilde{t}
\partial_{c_2},\\\\
\partial_{t_x}&=&\displaystyle \frac{\partial_{\tilde{t}}}{\partial_{t_x}}\frac{\partial}{\partial_{\tilde{t}}}
+\frac{\partial_{c_1}}{\partial_{t_x}}\frac{\partial}{\partial_{c_1}}
+\frac{\partial_{c_2}}{\partial_{t_x}}\frac{\partial}{\partial_{c_2}}
=-\partial_{c_1}+\beta \sinh \tilde{t}
\partial_{c_2},\\\\
\partial_{t_{xx}}&=&\displaystyle \frac{\partial_{\tilde{t}}}{\partial_{t_{xx}}}\frac{\partial}{\partial_{\tilde{t}}}
+\frac{\partial_{c_1}}{\partial_{t_{xx}}}\frac{\partial}{\partial_{c_1}}
+\frac{\partial_{c_2}}{\partial_{t_{xx}}}\frac{\partial}{\partial_{c_2}}=-\partial_{c_2}.
\end{eqnarray*}
\noindent Under these transformations $Y_{-1}$ becomes
\begin{equation}
Y_{-1}=\partial_{\tilde{t}}+\beta(\sinh \tilde{t}+\cosh
\tilde{t})\partial_{c_1}-\beta c_1(\sinh \tilde{t}+\cosh
\tilde{t})\partial_{c_2}.
\end{equation}
\noindent Again by using the method of characteristic,
\begin{equation}
\frac{d\tilde{t}}{1}=\frac{dc_1}{\beta(\sinh \tilde{t}+\cosh
\tilde{t})}=\frac{dc_2}{-\beta c_1(\sinh \tilde{t}+\cosh
\tilde{t})}=\frac{dI}{0}
\end{equation}
\noindent we obtain
\begin{equation}
\beta(\sinh \tilde{t}+\cosh \tilde{t})-c_1=\tilde{c}_1
\end{equation}
\noindent and
\begin{equation}
\frac{c_1^2}{2}+c_2=\tilde{c}_2.
\end{equation}
\noindent Thus from $Y_{-1}I=0$ we obtain the solution
\begin{equation}
I=G(\beta(\sinh \tilde{t}+\cosh \tilde{t})-c_1,\displaystyle \frac{c_1^2}{2}+c_2).
\end{equation}
But we search for the common solution of $Y_1I=0$ and $Y_{-1}I=0$. Since
the solution of $Y_1I=0$ only depends on $c_1$ and $c_2$ we take the solution
$I=H(u)$ where $u=\displaystyle \frac{c_1^2}{2}+c_2$. Note that we may simply take
\begin{equation}
I=\displaystyle \frac{c_1^2}{2}+c_2=\frac{\beta^2}{2}\cosh^2 t
-\beta t_x\cosh t+\frac{t_x^2}{2}+\beta t_x\sinh t-t_{xx}.
\end{equation}
\begin{rem}
We should also check that $DI=I$ to be sure that $I$ is
$n$-integral.
\begin{eqnarray*}
DI&=&\displaystyle
 \frac{\lambda^2}{2}t_x^2+\frac{\lambda^2\beta^2}{2}\cosh^2t
-\frac{\lambda^2\beta^2}{2}+\frac{\gamma^2}{2}+\lambda^2\beta t_x
\sinh t\\\\&+&\lambda \gamma t_x+\lambda \gamma \sinh t
 -\lambda t_{xx}-\lambda \beta t_x \cosh t.
\end{eqnarray*}
\noindent To make $DI$ similar to $I$, we have to choose
$\lambda=1$ and $\gamma=0$. In this case,
\begin{equation}
DI=I-\displaystyle  \frac{\beta^2}{2}.
\end{equation}
So $I$ is not $n$-integral, but it helps to find $n$-integral.
Really,
\begin{equation}
F=\frac{\beta^2}{2}\cosh^2 t -\beta t_x \cosh
t+\frac{t_x^2}{2}+\beta t_x \sinh t-t_{xx}+\frac{\beta^2}{2}n,
\end{equation}
solves the equation $DF=F$.
\end{rem}

\vspace{10mm}

\noindent \textbf{Problem 2.} The equation for the function $f$ is
\begin{equation}\label{problem2f}
t_{1x}=f(t,t_1,t_x)=\lambda (t_x+e^t)+e^{t_1}+c.
\end{equation}
Hence
\begin{equation}\label{problem2g}
t_{-1x}=g(t,t_x,t_{-1})=\frac{1}{\lambda} (t_x-e^t+\lambda e^{t_{-1}}-c).
\end{equation}
\noindent To obtain the operator $Y_1$ we find the equations
written below:
\begin{eqnarray*}
f_{t_1}&=&e^{t_1},\\\\
f_x &=& \lambda t_{xx}+\lambda t_x e^{t}+t_{1x}e^{t_1},\\\\
f_{xt_1}&=&t_{1x}e^{t_1}+e^{2t_1},\\\\
f_{xx}&=&\lambda t_{xxx}+\lambda (t_x)^2e^t+\lambda t_{xx}e^t
      +(\lambda t_{xx}+\lambda
      t_xe^t+t_{1x}e^{t_1})e^{t_1}+(t_{1x})^2e^{t_1},\\\\
f_{xxt_1}&=&e^{t_1}(\lambda t_{xx}+\lambda
t_xe^t+t_{1x}e^{t_1})+3t_{1x}e^{2t_1} +e^{3t_1}+(t_{1x})^2e^{t_1},
\end{eqnarray*}
\begin{eqnarray*}
f_{xxx}&=&\lambda t_{xxx}+2\lambda t_xt_{xx}e^t+\lambda
t_x^3e^t+\lambda t_x t_{xx}e^t
\\\\&+&3t_{1x}(\lambda t_{xx}+\lambda t_xe^t+t_{1x}e^{t_1})+(t_{1x}^3)e^{t_1}\\\\
&+&e^{t_1}\{\lambda t_{xxx}+\lambda t_{xx}e^t+\lambda t_x^2e^t
+[\lambda t_{xx}+\lambda t_xe^t+t_{1x}e^{t_1}]e^{t_1}+(t_{1x}^2)e^{t_1}\},\\\\
f_{xxxt_1}&=&(\lambda t_{xx}+\lambda t_xe^t+t_{1x}e^{t_1})(3t_{1x}e^{t_1}+3e^{2t_1})
+e^{t_1}t_{1x}(6t_{1x}e^{t_1}+3e^{2t_1})\\\\
&+&e^{t_1}\{\lambda t_{xxx}+\lambda t_{xx}e^t+\lambda t_x^2e^t
+[\lambda t_{xx}+\lambda t_xe^t+t_{1x}e^{t_1}]e^{t_1}+(t_{1x}^2)e^{t_1}\}\\\\
&+&e^{t_1}\{[\lambda t_{xx}+\lambda
t_xe^t+t_{1x}e^{t_1}]e^{t_1}+3e^{2t_1}t_{1x}+e^{3t_1}
+(t_{1x})^2e^{t_1}\}+(t_{1x})^3e^{t_1},\\\\
&\vdots&
\end{eqnarray*}
Hence
\begin{eqnarray*}
D^{-1}(f_{t_1})&=&e^t,\\\\
D^{-1}(f_{xt_1})&=&t_xe^t+e^{2t},\\\\
D^{-1}(f_{xxt_1})&=&t_{xx}e^t+3t_xe^{2t}+e^{3t}+t_x^2e^t,\\\\
D^{-1}(f_{xxxt_1})&=&3t_xt_{xx}e^t+7t_x^2e^{2t}
+4t_{xx}e^{2t}+6t_xe^{3t}+t_x^3e^t+t_{xxx}e^t+e^{4t},\\\\
&\vdots&
\end{eqnarray*}
\noindent Now we are ready to write the operator $Y_1$:
\begin{eqnarray}
Y_1&=&\partial_t+e^t\partial_{t_x}+(e^tt_x+e^{2t})\partial_{t_{xx}}
+(t_{xx}e^t+3t_xe^{2t}+e^{3t}+t_x^2e^t)\partial_{t_{xxx}}\nonumber\\\nonumber\\
&+&(3t_xt_{xx}e^t+7t_x^2e^{2t}
+4t_{xx}e^{2t}+6t_xe^{3t}+t_x^3e^t+t_{xxx}e^t+e^{4t})\partial_{t_{xxxx}}\nonumber\\\nonumber\\
&+&...
\end{eqnarray}
\noindent If we define $t_x+e^{t}=\psi$, $Y_1$ becomes

\begin{equation}
Y_1=\partial_t+e^t\partial_{t_x}+e^t\psi \partial_{t_{xx}}
+e^t(\psi^2+\psi_x)\partial_{t_{xxx}}
+e^t[(\psi^2+\psi_x)_x+\psi(\psi^2+\psi_x)]\partial_{t_{xxxx}}+...\,.
\end{equation}

\noindent Now we will obtain the operator $Y_{-1}$. From
(\ref{problem2g}) we have,
\begin{eqnarray*}
g_{t_{-1}}&=&-e^{t_{-1}},\\\\
g_x &=& [t_{xx}-t_x e^{t}-\lambda t_{-1x}e^{t_{-1}}]/\lambda,\\\\
g_{xt_{-1}}&=&-t_{-1x}e^{t_{-1}}+e^{2t_{-1}},\\\\
g_{xx}&=&[t_{xxx}-t_x^2e^t-t_{xx}e^t
      -(t_{xx}-t_xe^t-t_{-1x}e^{t_{-1}})e^{t_{-1}}-\lambda(t_{-1x})^2e^{t_{-1}}]/\lambda,
\end{eqnarray*}
\begin{eqnarray*}
g_{xxt_{-1}}&=&e^{t_{-1}}[-(t_{xx}- t_xe^t-\lambda t_{-1x}e^{t_{-1}})
+3\lambda t_{-1x}e^{t_{-1}}
-\lambda e^{2t_{-1}}-\lambda(t_{-1x})^2]/\lambda,\\\\
g_{xxx}&=&[t_{xxx}-3t_xt_{xx}e^t-t_x^3e^t
+3t_{-1x}e^{t_{-1}}(t_{xx}-t_xe^t-\lambda t_{-1x}e^{t_{-1}})-\lambda (t_{-1x})^3e^{t_{-1}}\\\\
&-&e^{t_{-1}}\{t_{xxx}-t_{xx}e^t-t_x^2e^t -(t_{xx}-t_xe^t-\lambda
t_{-1x}e^{t_{-1}})e^{t_{-1}}-\lambda(t_{-1x}^2)e^{t_{-1}}\}]/\lambda,
\end{eqnarray*}
\begin{eqnarray*}
g_{xxxt_{-1}}&=&[\lambda e^{t_{-1}}t_{-1x}(7t_{-1x}e^{t_{-1}}-6e^{2t_{-1}})\\\\
&+&(t_{xx}-t_xe^t-\lambda
t_{-1x}e^{t_{-1}})(-3t_{-1x}e^{t_{-1}}+4e^{2t_{-1}})\\\\
&-&e^{t_{-1}}\{t_{xxx}-t_{xx}e^t-t_x^2e^t
-(t_{xx}-t_xe^t-\lambda t_{-1x}e^{t_{-1}})e^{t_{-1}}-\lambda(t_{-1x}^2)e^{t_{-1}}\}\\\\
&+&e^{2t_{-1}}(t_{xx}-t_xe^t-\lambda
t_{-1x}e^{t_{-1}})e^{t_1}+\lambda
e^{4t_{-1}}-\lambda(t_{-1x})^3e^{t_{-1}}]/\lambda\\\\
&\vdots&
\end{eqnarray*}
From these equations we get
\begin{eqnarray*}
D(g_{t_{-1}})&=&-e^t,\\\\
D(g_{xt_{-1}})&=&e^{2t}-t_xe^t,\\\\
D(g_{xxt_{-1}})&=&-e^tt_{xx}+3t_xe^{2t}-e^{3t}-(t_x)^2e^t, \\\\
D(g_{xxxt_{-1}})&=&-3t_xt_{xx}e^t+7t_x^2e^{2t}
+4t_{xx}e^{2t}-6t_xe^{3t}-t_x^3e^t-t_{xxx}e^t+e^{4t},\\\\
&\vdots&
\end{eqnarray*}
\noindent Thus the operator $Y_{-1}$ is
\begin{eqnarray}
Y_{-1}&=&\partial_t-e^t\partial_{t_x}-(e^tt_x-e^{2t})\partial_{t_{xx}}
+(-e^tt_{xx}+3t_xe^{2t}-e^{3t}-(t_x)^2e^t)\partial_{t_{xxx}}\nonumber\\\nonumber\\
&+&(-3t_xt_{xx}e^t+7t_x^2e^{2t}
+4t_{xx}e^{2t}-6t_xe^{3t}-t_x^3e^t-t_{xxx}e^t+e^{4t})\partial_{t_{xxxx}}\nonumber\\\nonumber\\
&+&...\,.
\end{eqnarray}
\noindent If we define $t_x-e^{t}=\varphi$, $Y_{-1}$ becomes
\begin{equation}
Y_{-1}=\partial_t-e^t\partial_{t_x}-e^t\varphi \partial_{t_{xx}}
-e^t(\varphi^2+\varphi_x)\partial_{t_{xxx}}
-e^t[(\varphi^2+\varphi_x)_x+\varphi(\varphi^2+\varphi_x)]\partial_{t_{xxxx}}+...\,.
\end{equation}
\noindent Now we will only take the first three terms of $Y_1$ and
$Y_{-1}$ and find the integral $I$ such that both $Y_1 I=0$ and
$Y_{-1}I=0$ are satisfied. The first three terms of $Y_1$ are
\begin{equation}
Y_1=\partial_t+e^t\partial_{t_x}+(e^{2t}+t_xe^t)\partial_{t_{xx}}.
\end{equation}
\noindent The method of characteristic gives us
\begin{equation}
\displaystyle
\frac{dt}{1}=\frac{dt_x}{e^t}=\frac{dt_{xx}}{e^{2t}+t_xe^t}=\frac{dI}{0}.
\end{equation}
\noindent By the first and second terms we have
\begin{equation}
e^t-t_x=c_1.
\end{equation}
\noindent By the second and third terms we have
\begin{equation}
t_xe^t-t_{xx}=c_2.
\end{equation}
\noindent \noindent Hence from $Y_1I=0$ we get the solution $I=F(c_1,c_2)$ where
$F$ is an arbitrary function. Now by using the transformation $\tilde{t}=t$,
$c_1=e^t-t_x$, $c_2=t_xe^t-t_{xx}$, we will write $Y_{-1}$
 in terms of $\tilde{t}$, $c_1$ and $c_2$. Note that
\begin{eqnarray*}
\displaystyle
\partial_t&=&\frac{\partial_{\tilde{t}}}{\partial_t}\frac{\partial}{\partial_{\tilde{t}}}
+\frac{\partial_{c_1}}{\partial_t}\frac{\partial}{\partial_{c_1}}
+\frac{\partial_{c_2}}{\partial_t}\frac{\partial}{\partial_{c_2}}=
\partial_{\tilde{t}}+e^{\tilde{t}}\partial_{c_1}+t_xe^{\tilde{t}}\partial_{c_2},\\\\
\partial_{t_x}&=&\frac{\partial_{\tilde{t}}}{\partial_{t_x}}\frac{\partial}{\partial_{\tilde{t}}}
+\frac{\partial_{c_1}}{\partial_{t_x}}\frac{\partial}{\partial_{c_1}}
+\frac{\partial_{c_2}}{\partial_{t_x}}\frac{\partial}{\partial_{c_2}}=-\partial_{c_1}+e^{\tilde{t}}
\partial_{c_2},\\\\
\partial_{t_{xx}}&=&\frac{\partial_{\tilde{t}}}{\partial_{t_{xx}}}\frac{\partial}{\partial_{\tilde{t}}}
+\frac{\partial_{c_1}}{\partial_{t_{xx}}}\frac{\partial}{\partial_{c_1}}
+\frac{\partial_{c_2}}{\partial_{t_{xx}}}\frac{\partial}{\partial_{c_2}}=-\partial_{c_2}.
\end{eqnarray*}
\noindent Under these transformations $Y_{-1}$ becomes
\begin{equation}
Y_{-1}=\partial_{\tilde{t}}+2e^{\tilde{t}}\partial_{c_1}-2c_1e^{\tilde{t}}\partial_{c_2}.
\end{equation}
\noindent Again by using the method of characteristic,
\begin{equation}
\frac{d\tilde{t}}{1}=\frac{dc_1}{2e^{\tilde{t}}}=\frac{dc_2}{-2c_1e^{\tilde{t}}}=\frac{dI}{0}
\end{equation}
\noindent we obtain
\begin{equation}
2e^{\tilde{t}}-c_1=\tilde{c}_1
\end{equation}
\noindent and
\begin{equation}
\frac{c_1^2}{2}+c_2=\tilde{c}_2.
\end{equation}
\noindent Hence from $Y_{-1}I=0$ we obtain the solution
\begin{equation}
I=G(2e^{\tilde{t}}-c_1,\displaystyle \frac{c_1^2}{2}+c_2).
\end{equation}
But we search for the common solution of $Y_1I=0$ and $Y_{-1}I=0$. Since
the solution of $Y_1I=0$ only depends on $c_1$ and $c_2$ we take the solution
$I=H(u)$ where $u=\displaystyle \frac{c_1^2}{2}+c_2$. Note that we may simply take
\begin{equation}
I=\frac{c_1^2}{2}+c_2=\frac{e^{2t}}{2}+\frac{t_x^2}{2}-t_{xx}.
\end{equation}
\begin{rem}
We should also check that $DI=I$.
\begin{eqnarray*}
DI=\displaystyle
\frac{\lambda^2}{2}t_x^2+\frac{\lambda^2}{2}e^{2t} +\frac{c^2}{2}
+\lambda^2 t_x e^t+\lambda c t_x +\lambda c e^t-\lambda t_{xx}
-\lambda t_x e^t.
\end{eqnarray*}
\noindent  If we choose $\lambda=1$ and $c=0$, we see that $I$
becomes
\begin{equation}
I=\frac{e^{2t}}{2}+\frac{t_x^2}{2}-t_{xx}
\end{equation}
and it satisfies $DI=I$ i.e. $I$ is integral.
\end{rem}
\noindent Now we will find the $x$-integral of (\ref{problem2f})
 with the above conditions. Hence we try to find $F$ which
 satisfies,
\begin{equation}\label{prob3F}
D_x F(t_2,t_1,t)=F_{t_2}t_{2x}+F_{t_1}t_{1x}+F_tt_x=0.
\end{equation}
Here $t_{1x}=t_x+e^t+e^{t_1}$ and
$t_{2x}=t_{1x}+e^{t_1}+e^{t_2}=t_x+e^t+2e^{t_1}+e^{t_2}$. We
insert these into (\ref{prob3F}), we get
\begin{equation}
F_{t_2}(t_x+e^t+2e^{t_1}+e^{t_2})+F_{t_1}(t_x+e^t+e^{t_1})+F_tt_x=0.
\end{equation}
To satisfy this equation the coefficient of $t_x$ and the other
terms should vanish separately. Thus we have
\begin{eqnarray*}
&&F_{t_2}+F_{t_1}+F_t=0,\\
&&F_{t_2}(e^t+2e^{t_1}+e^{t_2})+F_{t_1}(e^t+e^{t_1})=0.
\end{eqnarray*}
So we get the operators,
\begin{eqnarray*}
&&X_1=\partial_{t_2}+\partial_{t_1}+\partial_t,\\
&&X_2=(e^t+2e^{t_1}+e^{t_2})\partial_{t_2}+(e^t+e^{t_1})\partial_{t_1}=0.
\end{eqnarray*}
The commutator of these operators is $[X_1,X_2]=X_2$.\\

\noindent Now for simplicity, we may consider the operator
$\tilde{X}_2=e^{-t}X_2$. Explicitly,
\begin{equation}
\tilde{X}_2=(1+e^{t_1-t})\partial_{t_1}+(1+2e^{t_1-t}+e^{t_2-t})\partial_{t_2}.
\end{equation}
\noindent We will make change of variables,
\begin{eqnarray*}
\tau&=&t,\\
\tau_1&=&t_1-t,\\
\tau_2&=&t_2-t.
\end{eqnarray*}
\noindent The operator $\tilde{X}_2$ becomes
\begin{equation}
\tilde{X}_2=(1+e^{\tau_1})\partial_{\tau_1}+(1+2e^{\tau_1}+e^{\tau_2})\partial_{\tau_2}.
\end{equation}
\noindent Now we use the method of characteristics;
\begin{equation}
\displaystyle
\frac{d\tau_1}{1+e^{\tau_1}}=\frac{d\tau_2}{1+2e^{\tau_1}+e^{\tau_2}}.
\end{equation}
Equivalently we have
\begin{equation}
\displaystyle
\frac{-de^{-\tau_1}}{1+e^{\tau_1}}=\frac{-de^{-\tau_2}}{1+2e^{\tau_1}+e^{\tau_2}}.
\end{equation}
\noindent Let $e^{-\tau_1}=u$ and $e^{-\tau_2}=v$. Hence we get
\begin{equation}
\displaystyle \frac{du}{1+u}=\frac{dv}{v(1+\frac{2}{u})+1}
\end{equation}
which has the solution
\begin{equation}
\displaystyle v=\frac{u^2}{u+1}\Big[\tilde{c}-\frac{1}{u}\Big].
\end{equation}
\noindent If we put $e^{-\tau_1}=u$ and $e^{-\tau_2}=v$ in this
solution we find $\tilde{c}$ as
\begin{equation}
\displaystyle
\tilde{c}=e^{\tau_1}\Big[1+\frac{(e^{\tau_1}+1)}{e^{\tau_2}}\Big].
\end{equation}
Hence $F$ is
\begin{equation}
F=e^{t_1-t}+e^{2t_1-t_2-t}+e^{t_1-t_2}.
\end{equation}
Thus, $t_{1x}=t_x+e^t+e^{t_1}$ is a discrete analog of the
Liouville equation.

\vspace{10mm}

\noindent \textbf{Problem 3.} The equation for the function $f$ is
\begin{equation}\label{problem3f}
t_{1x}=f(t,t_1,t_x)=r(t_x)+\beta t_1.
\end{equation}
Hence
\begin{equation}\label{problem3g}
t_{-1x}=g(t,t_x,t_{-1})=r^{-1}(t_x-\beta t).
\end{equation}
\noindent To obtain the operator $Y_1$ we find the equations
written below
\begin{eqnarray*}
f_{t_1}&=&\beta,\\\\
f_x&=&[r(t_x)]_x+\beta(r(t_x)+\beta t_1),\\\\
f_{xt_1}&=&\beta^2,\\\\
f_{xx}&=&[r(t_x)]_{xx}+\beta[r(t_x)]_x+\beta^2(r(t_x)+\beta
t_1),\\\\
&\vdots&
\end{eqnarray*}
\noindent Clearly the operator $Y_1$ is
\begin{equation}
Y_1=\partial_t+\beta \partial_{t_x}+\beta^2
\partial_{t_{xx}}+\beta^3 \partial_{t_{xxx}}+\beta^4
\partial_{t_{xxxx}}+...\,.
\end{equation}
\noindent Since the function $g=r^{-1}(t_x-\beta t)$ does not
depend on $t_{-1}$, the operator $Y_{-1}$ is
\begin{equation}
Y_{-1}=\partial_t.
\end{equation}

\begin{rem}
Let us take the first three terms of $Y_1$. Clearly, the operators
$Y_1$ and $Y_{-1}$ commutes. From $Y_{-1}I=\partial_t I=0$ we see
that $I$ is independent of $t$. From $Y_1I=0$ we obtain $I$ as
\begin{equation}
I=f(t_{xx}-\beta t_x).
\end{equation}
Simply, we may  take $I=t_{xx}-\beta t_x$. Now we check if $DI=I$.
\begin{equation}
DI=t_{xx}r_{t_x}(t_x).
\end{equation}
\noindent If $\beta=0$ and $r_{t_x}(t_x)=1$ the equation $DI=I$ is
satisfied. In this case the equation (\ref{problem3f}) takes the
form $t_{1x}=t_x$ and $I=t_{xx}$ is its $n$-integral.
\end{rem}

\section*{Acknowledgments}
The authors thank Prof. M. G\"{u}rses for fruitful discussions.
One of the authors (AP) thanks the Scientific and Technological
Research Council of Turkey (TUB{\.{I}}TAK) and the other (IH)
thanks (TUB{\.{I}}TAK), the Integrated PhD. Program (BDP) and
grants RFBR $\#$ 05-01-00775 and RFBR $\#$ 06-01-92051-CE$\_$a for
partial financial support.


\begin{thebibliography}{EMG}
\bibitem[1]{Darboux}
G. Darboux, {\it Le\c{c}ons sur la th$\acute{e}$orie
g$\acute{e}$n$\acute{e}$rale des surfaces et les applications
geometriques du calcul infinitesimal,} T.2. Paris: Gautier-Villars
(1915).
\bibitem[2]{GrundlandVassiliou}
A. M. Grundland, P. Vassiliou, {\it Riemann double waves, Darboux
method and the Painlev$\acute{e}$ property. Proc. Conf.
Painlev$\acute{e}$ transcendents, their Asymptotics and Physical
Applications,} Eds. D. Levi, P. Winternitz, NATO Adv. Sci. Inst.
Ser. B Phys, \textbf{278}, 163-174, (1992).
\bibitem[3]{SokolovZhiber}
V. V. Sokolov, A. V. Zhiber, {\it On the Darboux integrable
hyperbolic equations,} Phys. Lett. A, \textbf{208}, no:4-6,
303-308 (1995).
\bibitem[4]{AndersonKamran}
I. M. Anderson, N. Kamran, {\it The variational bicomplex for
hyperbolic second-order scalar partial differential equations in
the plane,} Duke Math. J. , \textbf{87}, no:2, 265-319 (1997).
\bibitem[5]{ShabatYamilov}
A. B. Shabat, R. I. Yamilov, {\it Exponential systems of type I
and the Cartan matrices,} (In Russian), Preprint, Bashkirian
Branch of Academy of Science of the USSR, Ufa, (1981).
\bibitem[6]{LeznovSmirnovShabat}
A. N. Leznov, V. G. Smirnov, A. B. Shabat, {\it Group of inner
symmetries and integrability conditions for two-dimensional
dynamical systems,} Teoret. Mat. Fizika, \textbf{51}, no:1, 10-21
(1982).
\bibitem[7]{AdlerStartsev} V. E. Adler, S. Ya. Startsev,
{\it On discrete analogues of the Liouville equation,} Teoret.
Mat. Fizika, \textbf{121}, no:2, 271-284 (1999), (English
translation: Theoret. and Math. Physics, \textbf{121}, no:2,
1484-1495, (1999)).
\bibitem[8]{NijhoffCapel}
F. W. Nijhoff, H. W. Capel, {\it The discrete Korteweg-de Vries
equation,} Acta Applicandae Mathematicae, \textbf{39}, 133-158
(1995).
\bibitem[9]{GKP}
B. Grammaticos, G. Karra, V. Papageorgiou, A. Ramani, {\it
Integrability of discrete-time systems, Chaotic dynamics,}
(Patras,1991), NATO Adv. Sci. Inst. Ser. B Phys, \textbf{298},
75-90, Plenum, New York, (1992).
\bibitem[10]{Habibullin}
I. T. Habibullin, {\it Characteristic algebras of fully discrete
hyperbolic type equations,} Symmetry, Integrability and Geometry:
Methods and Applications, no:1, paper 023, 9 pages, (2005) //
$\tt{arxiv: nlin.SI/0506027, 2005}$.
\bibitem[11]{IbragimovShabat}N. Kh. Ibragimov, A. B. Shabat,
{\it Evolution equations with nontrivial Lie-B\"{a}cklund group,}
Funktsional. Anal. i Prilozhen, \textbf{14}, no:1, 25-36 (1980).
\bibitem[12]{MSY} A. V. Mikhailov, A. B. Shabat, R. I. Yamilov,
{\it A symmetry approach to the classification of nonlinear
equations. Complete list of integrable systems,}(In Russian),
Uspekhi Mat. Nauk,  \textbf{42}, no:4, 3-53 (1987).
\bibitem[13]{LeviYamilov}
R. I. Yamilov, D. Levi, {\it Integrability conditions for $\$n\$$
and $\$t\$$ dependent dynamical lattice equations,} J. Nonlinear
Math. Phys. , \textbf{11}, no:1, 75-101 (2004).
\bibitem[14]{Yamilov}
R. I. Yamilov, {\it On classification of discrete evolution
equations,} Uspekhi Mat. Nauk, \textbf{38}, no:6, 155-156 (1983).
\bibitem[15]{AdlerBobenkoSuris}
V. E. Adler, A. I. Bobenko, Yu. B. Suris, {\it Classification of
integrable equations on quad-graphs. The consistency approach,}
Communications in Mathematical Physics, \textbf{233}, no:3,
513-543 (2003).
\bibitem[16]{Gurses1}
M. G\"{u}rses, A. Karasu, {\it Variable coefficient third order
KdV type of equations,} Journal of Math. Phys., \textbf{36}, 3485
(1995) // $\tt{arxiv: solv-int/9411004}$.
\bibitem[17]{Gurses2}
M. G\"{u}rses, A. Karasu, {\it Degenarate Svinolupov KdV Systems,}
Physics Letters A, \textbf{214}, 21-26 (1996).
\bibitem[18]{Gurses3}
M. G\"{u}rses, A. Karasu, {\it Integrable KdV Systems: Recursion
Operators of Degree Four,} Physics Letters A, \textbf{251},
247-249 (1999) // $\tt{arxiv: solv-int/9811013}$.
\bibitem[19]{Gurses4}
M. G\"{u}rses, A. Karasu, R. Turhan, {\it Nonautonomous Svinolupov
Jordan KdV Systems,} Journal of
 Physics A: Mathematical and General, \textbf{34}, 5705-5711 (2001) // $\tt{arxiv: nlin.SI/0101031}$.
\end{thebibliography}
\end{document}